\RequirePackage{fix-cm}
\documentclass[smallextended]{svjour3}       
\smartqed  
\usepackage{amsmath}
\usepackage{graphicx}
\usepackage{mathptmx}      
\begin{document}

\title{Distributing CV entanglement over 4 co-propagating orthogonal modes}

\author{Adriana Pecoraro         \and
        Alberto Porzio}

\institute{A. Pecoraro \at
              CNR - SPIN, Napoli \\
              Complesso Univ. Monte Sant'Angelo\\
              Via Cintia, 80126 Napoli, Italy\\
\email{pecoraro@fisica.unina.it}
           \and
           A. Porzio \at
              CNR - SPIN, Napoli \\
              Complesso Univ. Monte Sant'Angelo\\
              Via Cintia, 80126 Napoli, Italy\\
\email{alberto.porzio@spin.cnr.it}              }

\date{Received: date / Accepted: date}

\maketitle

\begin{abstract}
We propose a scheme for distributing continuous variable entanglement originally established among
a pair of mode between a set of four orthogonal co-propagating modes. This is accomplished
by exploiting the possibility of coupling polarization with optical angular momentum provided
by the \textit{q-plate}. Here we present the principle of the proposed scheme with a short feasibility study
that shows that the four-modes covariance matrix at the scheme output represent an entangled
multi mode system. 

\end{abstract}

\section{Introduction}
\label{intro}
One of the most peculiar traits of quantum systems is the presence
of correlations that cannot be explained by classical laws
of physics. In continuous variable (CV) quantum optics this translates into squeezing for single field mode \cite{Wu86}
and entanglement between distinct modes \cite{Ou92}.

Entanglement, firstly introduced by Schr\"odinger \cite{Schrodinger1935} in response to the famous
Einstein, Podolsky, and Rosen (EPR) paper in 1935 \cite{EPR}, plays a leading role
for what concerns applications indeed, together with coherence and
superposition, they are fundamental as quantum technology as emerged as the
strategy
to find enhanced ways of manipulating and transmitting information
\cite{QuantumRev} or to
overcome the classical limits in measurements \cite{Quantum Metrology}.

Here we will deal with CV entanglement in non-zero
Orbital Angular Momentum (OAM) beams. OAM represents a discrete quantum variable (DV) spanning an
infinite dimensional Hilbert space.
Mixing CV systems with DV features allows 
achieving quantum tasks not accessible with either photon-number states
or CV entanglement \cite{VanLoock2011,Andersen2015}.

In Ref. \cite{Pecoraro} we reported on the preparation and the complete
experimental characterization of a bipartite two-mode CV Gaussian
entangled state carrying OAM.

In the present paper, we propose a scheme for distributing the entanglement, originally cast
among a pair of modes in a type--II OPO, among a set of four mutually orthogonal modes.
This is possible by exploiting the higher dimensionality
provided by OAM.
As a matter of fact, the use of
a \textit{q-plate}, a liquid crystal device that couples polarization with
OAM \cite{Marrucci2006}, can realise a quantum beam-splitter among
co-propagating modes that acquire multi-distinguishability thanks to the twofold label
(polarization and OAM). In this way it is possible to distribute entanglement between
co-propagating modes that are mutually orthogonal.
The so obtained 4--modes Gaussian state is described by
a 8x8 covariance matrix in phase space.
Analysing the full system covariance matrix it can be found that
a single \textit{q-plate} allows to distribute entanglement
between the four output modes creating three possible combination
of pairs of entangled modes.
This happens at the expense of introducing fictitious losses
at the single mode level while preserving the total energy of the system.

The paper is structured as follows. In Sect. \ref{q-plate} we describe
the action of a \textit{q-plate} onto a pair of orthogonally polarised modes.
Then, in Sect. \ref{OPO}, the manipulation of the entangled state at the output of a type--II non-degenerate OPO is detailed. Section \ref{fourmode} discusses the properties of the
set of four modes at the \textit{q-plate} output while conclusions are, eventually, drawn in Sect. \ref{Concl}.  

\section{The \textit{q-plate} action}
\label{q-plate}

The \textit{q-plate} is a liquid crystal device that couples polarization (bi-dimensional)
d.o.f. with the infinite Hilbert space of optical orbital angular momentum (OAM). 

A beam transmitted through a \textit{q-plate}
gains $2q$ quanta of OAM where the $q$ parameter is the topological charge of the device.
The \textit{q-plate} is composed of a thin liquid
crystal film sandwiched between two glasses. The device is 
birefringent where $\delta$ is the retardation between the two axes.
The retardation $\delta$ can be tuned by applying an alternate voltage to the liquid crystal cell.

OAM is an important property of light known since
classical electromagnetic theory. Light, besides carrying
energy, can also transport momentum in its linear and angular components.
In particular, by focusing on the angular part, it's possible to distinguish
between the so-called spin angular momentum (SAM), macroscopically
associated with polarization, and the orbital angular momentum (OAM)
associated to the helicity of the phase fronts.

Both \textit{classical} properties of light are kept all the way down
to the single photon level and so to the quantum regime.
They arise from solving the Helmholtz equation in paraxial
approximation unveiling intrinsic properties of single photons.

More in detail, solving the most general form of the Helmholtz
equation gives particular kinds of beams, such as Laguerre--Gauss ones,
that are characterized by an azimuthal phase factor $e^{im\varphi}$ winding around the propagation axes.
In this case, the beam acquires a peculiar helical wavefront, and shows, in
a transverse plane, a doughnut intensity profile with an optical vortex
at the centre due to the therein located phase singularity \cite{PadgettOAM,OAMRev}.

The value $\left|m\right|$
indicates that each photon in the beam carries $\left|m\right|$ quanta of OAM.
Since in principle $m$
can assume any positive and negative integer value, this internal d.o.f. of the photon
exploits an infinite dimensional
Hilbert space allowing the possibility of encoding a large amount
of informations onto a single beam.

Given at the \textit{q-plate} input a 
generic \textit{k} mode
of orbital order \textit{m}, its action, in the $L$ (left) and $R$ (right) circularly polarisation base, is:
\begin{align}
\hat Q\; k_{\left[L,m\right]}= \cos \frac{\delta}{2}\; k_{\left[L,m\right]} +i\sin \frac{\delta}{2}\; k_{\left[R,m+2q\right]} \nonumber \\
\hat Q\; k_{\left[R,m\right]} = \cos \frac{\delta}{2}\; k_{\left[R,m\right]} +i\sin \frac{\delta}{2}\; k_{\left[L,m-2q\right]}
\label{Eq:q-plate}
\end{align}
where $\hat{Q} $ represents the \textit{q-plate} operator,
and $\left[ A,m \right] $ labels the mode with OAM $m$ and polarization $A= (L, R)$.

From Eq. (\ref{Eq:q-plate}) it is easy to see that a \textit{q-plate}
couples polarization with OAM. In particular, entering the device with a circularly polarised
light beam and a well defined OAM order \textit{m}, gives raise to a pair of beams that are distinguishable for both d.o.f..
Tuning the birefringence retardation $\delta$
allows to change the weights of the output components and, in particular,
setting $\delta=\pi$ a single circularly polarised beam travelling across the \textit{q-plate}
reverses its polarization and acquires $\pm 2q$ quanta of OAM.
This regime has been exploited for endowing a cross polarised pair of CV
entangled beams with OAM extra--distinguishability through the
\textit{q-plate} action as reported in \cite{Pecoraro}.
 
Classically the \textit{q-plate} acts as a device with one input and two outputs.
From the quantum point of view this would be in contrast with the preservation of the Heisenberg uncertainty principle.
From the quantum perspective, the \textit{q-plate} is analogous to a beam--splitter acting onto
the SAM+OAM mode space. As it will be detailed in the next section it couples two input mode to two output modes. 
The quantum action of the \textit{q-plate} must take into account the \textit{vacuum} modes
that take part into the complete transformation.
In particular, tuning the retardation $\delta$ to $\pi/2$ realise a balanced beam splitter acting
on co-propagating orthogonal modes.

\section{Manipulating a bi--partite CV polarization entangled state}
\label{OPO}

Our starting
point is a CV Gaussian bipartite two-mode entangled state generated by
a type--II Optical Parametric Oscillator (OPO) working in frequency
degeneration \cite{DAuria08}. Such a device generates two entangled
frequency-degenerate continuous-wave beams exploiting the polarization
as the distinguishing degree of freedom (d.o.f) \cite{DAuria09}. These two
beams are in the fundamental a Gaussian mode ($TEM_{00}$),
have linear orthogonal polarization and are collinear.

Let's indicate these two modes with $a_{\left[H,0\right]}$ and $b_{\left[V,0\right]}$ ($\left[H,0\right]$ and $\left[V,0\right]$ are the polarization and the initial OAM value for the respective mode).

\begin{figure}
	\centering\includegraphics[width=0.88\textwidth]{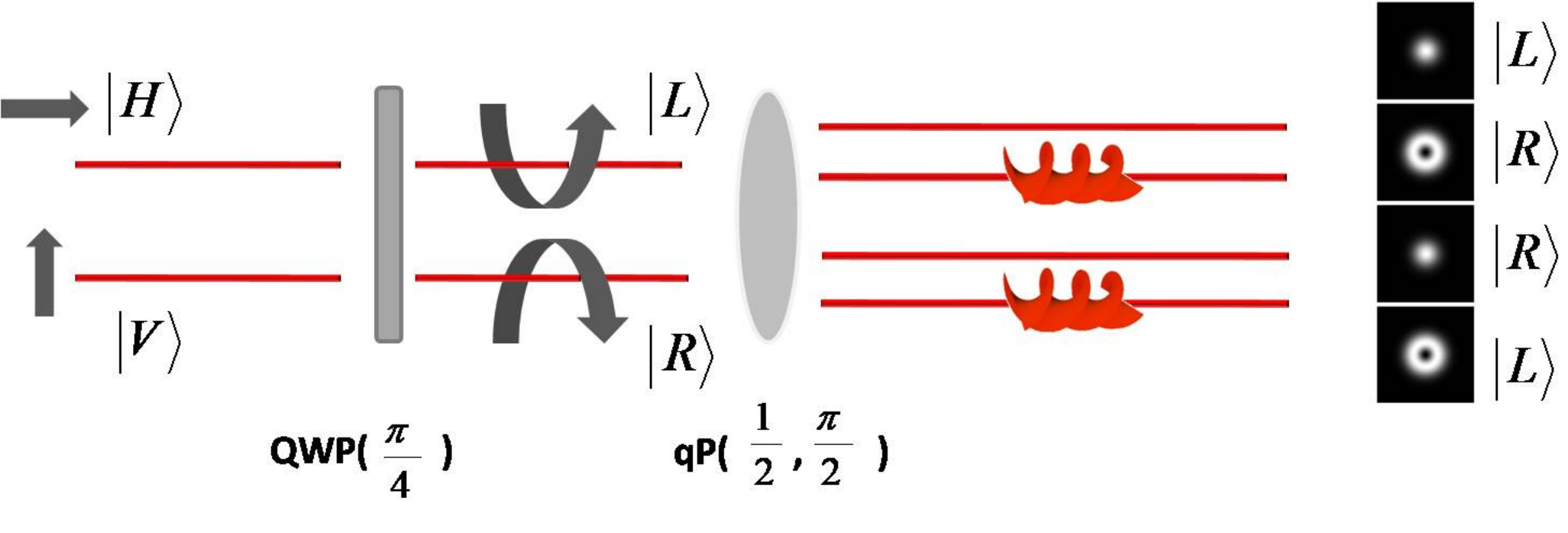}
	\caption{Scheme for the distribution of bipartite entanglement among four distinct orthogonal modes at the ouput of a type--II OPO. The initial orthogonally lineraly polarized pair pass through a quarter-waveplate for getting into an orthogonally circularly polarised pair. Then, passing through a properly tuned \textit{q-plate} allows to access four modes distinguishable by either polarization and/or optical OAM.}
	\label{Fig:scheme}       
\end{figure}

The scheme we are going to discuss in some details is sketched in Fig. \ref{Fig:scheme}.

The first step toward the distribution of the entanglement over four distinct modes is the use of a quarter-waveplate to makes the two OPO output beams circularly polarized so that $a_{\left[H,0\right]}\rightarrow a_{\left[L,0\right]}$ and  
 $b_{\left[V,0\right]}\rightarrow b_{\left[R,0\right]}$.

Then, the two circularly polarised beams enters a $q=1/2$ \textit{q-plate} tuned at $\delta=\pi/2$.
To have a quantum picture of the \textit{q-plate} induced transformation we have to consider a reversible version of Eq. (\ref{Eq:q-plate}).
In order to obtain these relations we have to consider a
transformation that involves, since the input, the 4 modes  with labels
$\left(\left[R,0\right],\left[L,0\right],\left[R,1\right],\left[L,-1\right]\right)$. The fully quantum version for Eq. (\ref{Eq:q-plate}) is
\begin{align}
\hat Q\; a_{\left[L,0\right]}= \frac{1}{\sqrt{2}}
\left(a_{\left[L,0\right]} +i\,a_{\left[R,1\right]}\right) \nonumber \\
\hat Q\; a_{\left[R,1\right]}= \frac{1}{\sqrt{2}}
\left(a_{\left[R,1\right]} +i\,a_{\left[L,0\right]}\right) \nonumber \\
\hat Q\; b_{\left[R,0\right]}= \frac{1}{\sqrt{2}}
\left(b_{\left[R,0\right]} +i\,b_{\left[L,-1\right]}\right) \nonumber \\
\hat Q\; b_{\left[L,-1\right]}= \frac{1}{\sqrt{2}}
\left(b_{\left[L,-1\right]} +i\,b_{\left[R,0\right]}\right) 
\label{Eq:quantum-q-plate}
\end{align}
It is straightforward to see that the above set of equations is made of two uncoupled sub-sets.
For distinguishing the mode \textit{in} and \textit{out} from the \textit{q-plate}, from now on we will identify the set of modes at the output by:
\begin{eqnarray}
a_{1} & \equiv & a_{[L,0]}\nonumber \\
a_{2} & \equiv & a_{[R,1]}\nonumber \\
b_{1} & \equiv & b_{[R,0]}\nonumber \\
b_{2} & \equiv & b_{[L,-1]}
\label{eq:modi-1-1}
\end{eqnarray}
Thanks to the introduction of OAM d.o.f. these modes are all orthogonal and so physically distinguishable.

In order to investigate the presence of mutual correlations among the set of output modes, we exploit the
fact that the state produced by the OPO source is Gaussian i.e. it
possesses a Gaussian Wigner function in phase space \cite{Olivares,Gaussian}.
Hence, thanks to the Gaussian character, that is preserved by linear transformations as 
(\ref{Eq:quantum-q-plate}), the state of the four modes system can be completely characterized by
its Covariance Matrix (CM) that, in this case, is a $8\times8$ symmetric positive-definite
matrix.
To construct the state CM let us introduce pair of canonical quadrature operators i.e. for each mode of the set:
\begin{align*}
X_{k_{i}}=\frac{k_{i}+k_{i}^{\dagger}}{\sqrt{2}}\\ Y_{k_{i}}=\frac{k_{i}-k_{i}^{\dagger}}{\sqrt{2}i}
\end{align*}
where $k=a,b$ and $i=1,2$. The second statistical moments of these two operators are the elements of the CM,
\begin{equation}
\Sigma_{4}=\left(\begin{array}{cccc}
\sigma_{a_{1}} & \varepsilon_{12} & \varepsilon_{13} & \varepsilon_{14}\\
\varepsilon_{12}^{T} & \sigma_{a_{2}} & \varepsilon_{23} & \varepsilon_{24}\\
\varepsilon_{13}^{T} & \varepsilon_{23}^{T} & \sigma_{b_{1}} & \varepsilon_{34}\\
\varepsilon_{14}^{T} & \varepsilon_{24}^{T} & \varepsilon_{34}^{T} & \sigma_{b_{2}}
\end{array}\right)\label{eq:sigma8}
\end{equation}
where the $\sigma$ blocks are single mode CM while $\varepsilon$ blocks contain information on the
possibly quantum mutual correlation between distinct modes. They are given by: 
\begin{eqnarray}
\sigma_{k_{i}}=\left(\begin{array}{cc}
\triangle X_{k_{i}}^{2} & \triangle X_{k_{i}}Y_{k_{i}}\\
\triangle Y_{k_{i}}X_{k_{i}} & \triangle Y_{k_{i}}^{2}
\end{array}\right) &  & \varepsilon_{ij}=\left(\begin{array}{cc}
\triangle X_{k_{i}}X_{l_{j}} & \triangle X_{k_{i}}Y_{l_{j}}\\
\triangle Y_{k_{i}}X_{l_{j}} & \triangle Y_{k_{i}}Y_{l_{j}}
\end{array}\right)
\end{eqnarray}
where $\triangle A_{k_{i}}^{2}$ are quadrature variances while covariance $\triangle A_{k_{i}}B_{l_{j}}$ are expressed as
\begin{equation*}
\triangle A_{k_{i}}B_{l_{j}} =\frac{1}{2}\left\langle \left\{ A_{k_{i}},B_{l_{j}}\right\} \right\rangle -\left\langle A_{k_{i}}\right\rangle \left\langle B_{l_{j}}\right\rangle 
\end{equation*}

The presence of entanglement among pair of modes can be witnessed by applying suitable criteria
to the elements of this matrix. A necessary and sufficient criterion
for the (non--)separability of bipartite Gaussian states with an arbitrary
number of modes has been established by Giedke et al. \cite{lewenstein}.
We will adopt this criterion later on to verify the presence of entanglement
between realistic covariance matrices obtained transforming a pair of modes at the OPO output.

\section{Properties of the four modes state}
\label{fourmode}

Let's now focus on the quadrature of the modes at the \textit{q-plate} output to properly construct the final CM with respect to the CM characterizing the pair of modes at the OPO output.
As seen the set \ref{Eq:quantum-q-plate} can be split in two  pairs descending from the OPO mode $a_{[H,0]}$ and $b_{[V,0]}$ respectively.

Inverting the first two Eqs. \ref{Eq:quantum-q-plate} one obtains
\begin{align}
a_{1}  =  \frac{1}{\sqrt{2}}\left(a_{[L,0]}-i\tilde{a}_{[R,1]}\right)\nonumber \\
a_{2}  =  \frac{1}{\sqrt{2}}\left(\tilde{a}_{[R,1]}-ia_{[L,0]}\right)\label{eq:trasfinvA}
\end{align}
So that the
relations among the quadratures of the two sides of the \textit{q-plate} are:
\begin{align}
X_{a_{1}}=\frac{1}{\sqrt{2}}\left(a_{1}+a_{1}^{\dagger}\right)=\frac{1}{\sqrt{2}}\left(X_{a_{[L,0]}}+\tilde{Y}_{a_{[R,1]}}\right)\nonumber \\
Y_{a_{1}}=\frac{1}{\sqrt{2}i}\left(a_{1}-a_{1}^{\dagger}\right)=\frac{1}{\sqrt{2}}\left(Y_{a_{[L,0]}}-\tilde{X}_{a_{[R,1]}}\right)\nonumber \\
X_{a_{2}}=\frac{1}{\sqrt{2}}\left(a_{2}+a_{2}^{\dagger}\right)=\frac{1}{\sqrt{2}}\left(\tilde{X}_{a_{[R,1]}}+Y_{a_{[L,0]}}\right)\nonumber \\
Y_{a_{2}}=\frac{1}{\sqrt{2}i}\left(a_{2}-a_{2}^{\dagger}\right)=\frac{1}{\sqrt{2}}\left(\tilde{Y}_{a_{[R,1]}}-X_{a_{[L,0]}}\right)
\end{align}
Similarly, for the pair of mode coming from $b_{[V,0]}$ one gets:
\begin{eqnarray}
b_{1} & = & \frac{1}{\sqrt{2}}\left(b_{[R,0]}-i\tilde{b}_{[L,-1]}\right)\nonumber \\
b_{2} & = & \frac{1}{\sqrt{2}}\left(-ib_{[R,0]}+\tilde{b}_{[L,-1]}\right)\label{eq:trasfinvB}
\end{eqnarray}
and the relations among quadratures are:
\begin{align}
X_{b_{1}}=\frac{1}{\sqrt{2}}\left(b_{1}+b_{1}^{\dagger}\right)=\frac{1}{\sqrt{2}}\left(X_{b_{[R,0]}}+\tilde{Y}_{b_{[L,-1]}}\right)\nonumber \\
Y_{b_{1}}=\frac{1}{\sqrt{2}i}\left(b_{1}-b_{1}^{\dagger}\right)=\frac{1}{\sqrt{2}}\left(Y_{b_{[R,0]}}-\tilde{X}_{b_{[L,-1]}}\right)\nonumber \\
X_{b_{2}}=\frac{1}{\sqrt{2}}\left(b_{2}+b_{2}^{\dagger}\right)=\frac{1}{\sqrt{2}}\left(Y_{b_{[R,0]}}+\tilde{X}_{b_{[L,-1]}}\right)\nonumber \\
Y_{b_{2}}=\frac{1}{\sqrt{2}i}\left(b_{2}-b_{2}^{\dagger}\right)=\frac{1}{\sqrt{2}}\left(\tilde{Y}_{b_{[L,-1]}}-X_{b_{[R,0]}}\right)
\end{align}
If the initial bipartite two-mode state (of the pair $\left(a_{[H,0]},b_{[V,0]}\right)$) is described
by a $2\times2$ CM in the standard form, i.e.: 
\begin{equation}
\Sigma_{2}=\left(\begin{array}{cccc}
a & 0 & c_{1} & 0\\
0 & a & 0 & c_{2}\\
c_{1} & 0 & b & 0\\
0 & c_{2} & 0 & b
\end{array}\right)
\end{equation}
one finds that $\Sigma_{4}$ can
be written in terms of the elements of $\sigma_{2}$ as
\begin{equation}
\Sigma_{4}=\frac{1}{2}\left(\begin{array}{cccccccc}
a+sn & 0 & 0 & sn-a & c_{1} & 0 & 0 & -c_{1}\\
0 & a+sn & a-sn & 0 & 0 & c_{2} & c_{2} & 0\\
0 & a-sn & a+sn & 0 & 0 & c_{2} & c_{2} & 0\\
sn-a & 0 & 0 & a+sn & -c_{1} & 0 & 0 & c_{1}\\
c_{1} & 0 & 0 & -c_{1} & b+sn & 0 & 0 & sn-b\\
0 & c_{2} & c_{2} & 0 & 0 & b+sn & b-sn & 0\\
0 & c_{2} & c_{2} & 0 & 0 & b-sn & b+sn & 0\\
-c_{1} & 0 & 0 & c_{1} & sn-b & 0 & 0 & b+sn
\end{array}\right)
\label{Eq:sigma4-th}
\end{equation}
where $sn$ refers to the shot noise of the vacuum modes $sn=\triangle\widetilde{X}_{b_{[L,-1]}}^{2}=\triangle\widetilde{Y}_{b_{[L,-1]}}^{2}=\triangle\widetilde{X}_{a_{[R,1]}}^{2}=\triangle\widetilde{Y}_{b_{[R,1]}}^{2}=1/2$.

To perform a realistic feasibility test we have applied the above transformations to the experimental matrix $\Sigma_{2}$ reported in Ref. \cite{Pecoraro} and corrected for collection losses.
The matrix so obtained represents the pure state as it is generated
inside the OPO crystal and it is given by:
\begin{equation}
\sigma_{exp}=\left(\begin{array}{cccc}
0.72\pm0.05 & 0 & 0.51\pm0.01 & 0\\
0 & 0.72\pm0.05 & 0 & -0.51\pm0.01\\
0.51\pm0.01 & 0 & 0.72\pm0.05 & 0\\
0 & -0.51\pm0.01 & 0 & 0.72\pm0.05
\end{array}\right)\label{eq:expPuro}
\end{equation}
Putting these values in the matrix (\ref{Eq:sigma4-th}) one obtains
\begin{equation}
\sigma_{4(exp)}=\left(\begin{array}{cccccccc}
0.60 & 0 & 0 & -0.10 & 0.25 & 0 & 0 & -0.25\\
0 & 0.60 & 0.10 & 0 & 0 & -0.25 & -0.25 & 0\\
0 & 0.10 & 0.60 & 0 & 0 & -0.25 & -0.25 & 0\\
-0.10 & 0 & 0.60 & 0.60 & -0.25 & 0 & 0 & 0.25\\
0.25 & 0 & 0 & -0.25 & 0.60 & 0 & 0 & -0.10\\
0 & -0.25 & -0.25 & 0 & 0 & 0.60 & 0.10 & 0\\
0 & -0.25 & -0.25 & 0 & 0 & 0.10 & 0.60 & 0\\
-0.25 & 0 & 0 & 0.25 & -0.10 & 0 & 0 & 0.60
\end{array}\right)\label{eq:guessed}
\end{equation}
We set a Mathematica routine to apply to the above matrix the the Giedke et al. iterative criterion proving the presence of entanglement
between pairs of subsystems. 
In particular, each mode has two entangled companions. $a_1$ is entangled with $b_1$ and $b_2$
(and vice-versa), the same happens for $a_2$.
On the contrary, $a_1$ and $a_2$ are separable and the same is for $b_1$ and $b_2$.

\section{Conclusions}
\label{Concl}

In conclusion we have proposed a scheme to distribute the entanglement
between two e.m. modes produced by a standard type-II phase matching
OPO source among four modes so realizing a bipartite four-modes entangled
state. This goal has been accomplished thanks to the introduction
of the OAM degree of freedom. Vortex modes have been obtained by a \textit{q-plate} that, by properly tuning its parameter $\delta$ , permits
to split each of the two initial entangled modes into two further
modes distinguishable by both OAM and polarization d.o.f. so accessing a larger Hilbert space.
The Gaussian
OAM-carrying bipartite four-modes state obtained in this way, can be
fully characterized
by its CM whose form, in terms of the initial pair of modes
has been calculated.
We have used an experimentally measured matrix in order to show that this scheme would effectively generate a bipartite four-modes entangled state.
The entanglement of the final matrix has been established applying criterion introduced by Giedke et al. \cite{lewenstein}

\end{document}